\documentclass[aps,prl,preprint]{article}

\usepackage{latexsym}
\usepackage{graphicx}

\title{Sensitive  boundary condition dependence of noise-sustained structure}
\author{Koichi Fujimoto\thanks{fujimoto@complex.c.u-tokyo.ac.jp}
 and Kunihiko Kaneko\\
  {\small \sl Department of Pure and Applied Sciences}\\
  {\small \sl University of Tokyo, Komaba, Meguro-ku, Tokyo 153, JAPAN}\\}
\date{submitted to Phys.Rev.Lett.}

\begin{document}
%\draft
\maketitle

\begin{abstract}
Sensitive boundary condition dependence (BCD) is reported
in a convectively unstable system with  noise, where
the amplitude of generated oscillatory dynamics in the downstream 
depends sensitively on the boundary value. 
This BCD is explained in terms of the manner in which
 the co-moving Lyapunov exponent
(characterizing the convective instability) decreases
from upstream to downstream.  It is shown that a fractal BCD
appears if the dynamics that represent the spatial change of the fixed point
includes transient chaotic dynamics.  By considering as an example a
one-way-coupled map lattice, this theory for BCD is demonstrated.
\end{abstract}

%\pacs{05.45.-a, 05.40.-a, 05.45.Ra}

Convective instability, which causes amplification of a disturbance 
along flow\cite{Lifshitz,Huerre}, 
is generally important in open flow systems \cite{Huerre,Deissler-1}.
If a system is convectively unstable (CU), a small disturbance at an
upstream position is amplified and transmitted downstream.
Due to this property, spatiotemporal structure with a
large amplitude can be generated  in the downstream by a tiny 
fluctuation in the upstream. 
Such structure is referred to as noise-sustained 
structure (NSS)\cite{Deissler-1}.
   
Convective instability is quantitatively characterized by a co-moving 
Lyapunov exponent $\lambda_v$, i.e., the 
Lyapunov exponent observed in an inertial system moving with the 
velocity $v$ \cite{KK-Deissler,KK-Co-move}. 
If $\mathop{max}_{v}\lambda_v$ is positive for a given state, 
the state is convectively unstable.
This condition is  compared to that for  linear instability,
implies $\lambda_0>0$. 
Absolute stability (AS), which implies stability along any flow\cite{Lifshitz,Huerre},
is guaranteed by the condition $\mathop{max}_{v}\lambda_v<0$. 
(With respect to an attractor, the co-moving Lyapunov exponent is used as an
indicator of chaos: chaos with convective instability is 
characterized by the positivity of $\mathop{max}_{v}\lambda_v$. 
However, with respect to a state, it is generally used to  
characterize its stability.)

In a system with convective instability and noise, we have shown that the 
 downstream dynamics depend on the boundary condition at the upstream.
Such behavior exhibiting some threshold-type dependence on the boundary 
condition was identified and analyzed in
connection with the change of the convective instability 
along the flow\cite{Fuji-KK}.
In the present letter, we demonstrate that {\sl sensitive}
BCD of the downstream dynamics can appear
in a class of noisy open-flow systems, and we clarify the condition for 
this appearance.

In this letter we consider the simple case of a discrete system in one spatial
dimension of the kind described above.
As a simple example, we adopt a one-way coupled map lattice (OCML)
\cite{KK-Co-move,Deissler-0,KK0,Frederick-KK,KK-Crutch,Vivaldi}  with noise: 
\begin{equation}
\label{eqn:CML}
x_n^i=(1-\epsilon)f(x_{n-1}^i) + \epsilon f(x_{n-1}^{i-1}) + \eta_n^i  .
\end{equation}
Here $n$ is a discrete time step, $i$ is the index denoting elements 
($i = 1,2,,,,M = $ system size), and $\eta$ is a white noise satisfying
$<<\eta_n^i \eta_m^j>> =\sigma^2 \delta_{nm} \delta_{ij}$, with $<<...>>$
representing an ensemble average ($\sigma \ll 1$)\cite{Comm0}.
A {\sl fixed boundary condition} $x^0$ is adopted and the (sensitive) dependence of the
downstream dynamics on $x^0$ is  studied.
The use of a CML here is just for convenience for illustration. 
The results and theory we present are straightforwardly
adapted to the  case of  coupled ordinary differential equations (ODE).

\begin{figure}[b]
\begin{center}
\includegraphics[scale=0.50]{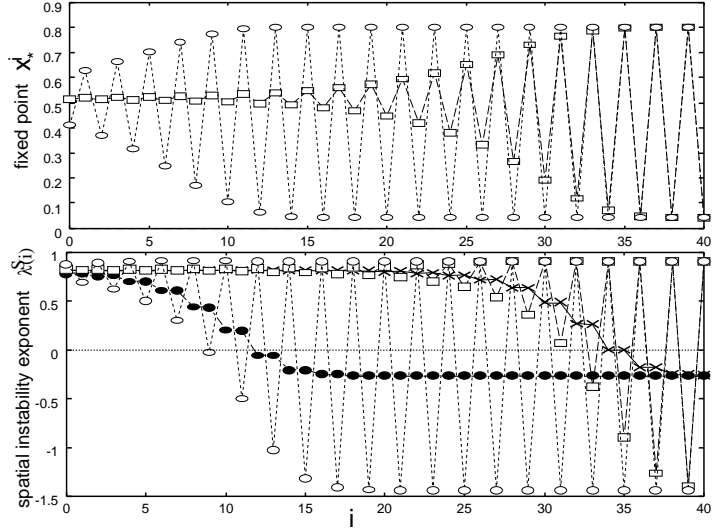}
\caption{The patterns $x_*^i$  and $\lambda^S(i)$.
The upper figure displays the fixed-point pattern $x_*^i$ in the absence of 
noise, 
while  the lower figure displays  the spatial pattern of $\lambda^S(i)$.
Two sets of plots are overlaid for different values of the boundary,
$x^0 = 0.52$ (solid line with $\Box$) and $0.60$ (dotted line with $\circ$).
For the lower figure $\sum_{\ell =0}^1 \lambda^S(i-\ell) / 2$ 
($\times$ for $x^0 = 0.52$, $\bullet$ for  $0.60$) 
is also plotted.
This difference reflects the difference in the downstream dynamics 
in the case with noise. 
For the boundary value for  $x^0 = 0.60$ no oscillation is generated  
in the downstream  (since the CU region is narrow, as discussed in the text), 
while oscillation is generated in the downstream  for 
$x^0 = 0.52$, as shown in Fig.3.
Here, $a = 1.8, \epsilon = 0.83$.}
\label{fig:i-fix}
\end{center}
\end{figure}

\begin{figure}
\begin{center}
\includegraphics[scale=0.42]{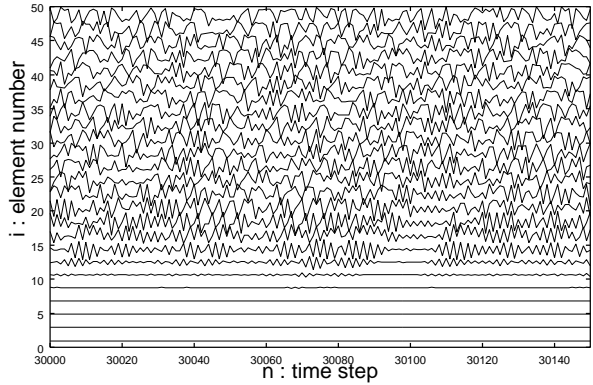}
\caption{Spatiotemporal plot of NSS with $\sigma = 2 \times 10^{-3}$, 
$a = 1.8, \epsilon = 0.83, x^0 = 0.52$.
Here $x_n^i$ is plotted for every second lattice site.}
\label{fig:i-x}
\end{center}
\end{figure}

In the present case we choose the logistic map $f(x)=1-a x^2 \ (-1 < x <1)$. 
The parameters $a$ and $\epsilon$ are chosen so that in the noiseless case 
all elements are attracted to fixed points $x_n^i=x_*^i$ for any 
initial and boundary conditions [i.e., the fixed points (in time) 
are AS in the  downstream].
This attraction to fixed points is realized in the  strong coupling regime 
(see Ref.\cite{Frederick-KK}). 
The values of the fixed points 
can depend on the lattice site number (i.e. can be functions of space),
and set of points \{$x_*^i$\} form a spatially periodic pattern,
as show in Fig.\ref{fig:i-fix}.
When noise is added, 
however, it can be amplified in the  downstream to create oscillating motion
(see Fig.\ref{fig:i-x}) if the upstream fixed points are convectively 
unstable \cite{Deissler-1}.

\begin{figure}
\begin{center}
\includegraphics[scale=0.48]{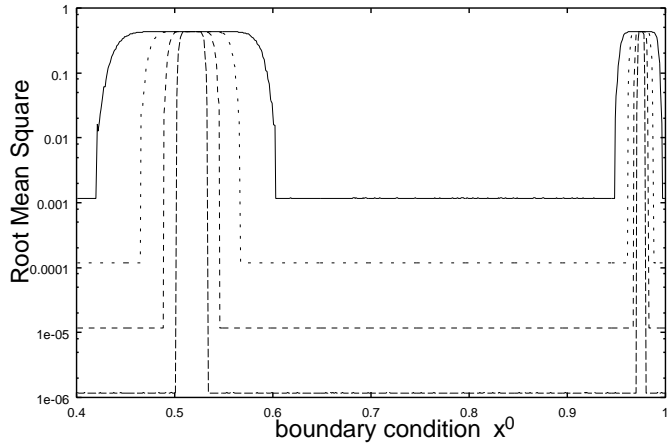}
\caption{BCD of the RMS for the downstream dynamics.
NSS appears around $x^0 = 0.52$ and  $0.97$. Except these two regions for the
boundary value, $x_n^i$ in the downstream falls on fixed points.
Different lines correspond to different noise intensities:
$\sigma = 2 \times 10^{-6},2 \times  10^{-5}, 2 \times 10^{-4}, 
2 \times 10^{-3}$.  
With the increase in the strength of the noise, 
the size of the region in which NSS is found increases.
Here $\epsilon = 0.842$, and the fixed point pattern is spatially period 2.}
\label{fig:i-bun}
\end{center}
\end{figure}

Whether or not this noise-sustained structure is formed in the downstream  
depends on the boundary value $x_0$ and the noise strength.
As a rough measure for the amplitude of the downstream oscillation, the root mean square (RMS)
$\sqrt{<x_n^i-<x_n^i>>^2}$ is computed, where $<...>$ is the temporal average.
As shown in Fig.\ref{fig:i-bun}, the RMS has a threshold-type
dependence on $x_0$.
Such a BCD has been observed in a one-way coupled ODE\cite{Fuji-KK}.  
The mechanism responsible for the sensitive BCD
clarified in that study  is universal
and can be summarized as follows:
 A change in  $x^0$ causes a change in the value of 
upstream fixed points $x_*^i$, which in turn causes a change in the degree of 
convective instability. 
 Accordingly the  downstream dynamics, generated through the spatial 
amplification of noise, can be different for different values of $x^0$. 

In the present case this mechanism is quantitatively analyzed as follows:
The spatial fixed points dynamics are described by the spatial recursive equation
\begin{equation}
\begin{array}{rcl}
\label{eq:spacemap}
x_*^i & = & (1-\epsilon)f(x_*^i) + \epsilon f(x_*^{i-1}) \\
& = & \frac{-1 + \sqrt{1+4 a(1-\epsilon)(1-a \epsilon (x_*^{i-1})^2)}}{2a(1-\epsilon)} \\
& \equiv & g(x_*^{i-1}) ,
\end{array}
\end{equation}
while the co-moving Lyapunov exponent $\lambda_v(i)$ of each fixed point 
$x_*^i$ \cite{Fuji-KK,KK-Crutch} is given by 
\begin{equation}
\begin{array}{rcl}
\label{eq:Co-move}
\lambda_v(i) & = &(1-\epsilon) \mbox{log} |f'(x_*^i)|  + \epsilon \mbox{log} |f'(x_*^{i-1})| \\
&  & + \mbox{log} |\frac{1 - \epsilon}{1 - v}| + \mbox{log}|\frac{\epsilon(1-v)}{v(1-\epsilon)}|^v .
\end{array}
\end{equation} 
A relevant quantity, characterizing the amplification of a disturbance per
lattice site is given by the
spatial instability exponent $\lambda^S(i)$ \cite{Vulpiani},
\begin{equation}
\label{eq:Sp-Lyap}
\lambda^S(i)=max_v \frac{\lambda_v(i)}{v} . 
\end{equation}
In Fig.\ref{fig:i-fix}, $\lambda^S(i)$ is also plotted as a function of 
the lattice site number. 
For both the boundary values $x^0 = 0.52$ ($\Box$) and $x^0 = 0.60$ ($\circ$),
convective instability exists in the upstream, 
as indicated by $\lambda^S(i)$.
(Here the downstream pattern has spatial period 2, and $\lambda^S(i)$ 
also oscillates with this period.
For this reason, $\sum_{\ell =0}^1\lambda^S(i-\ell) / 2$, the average over the 
spatial period, is also plotted.)
For  $x^0 = 0.60$, $\lambda^S(i)$ (or the average over the spatial period)
becomes negative at a smaller lattice site number than the case
for $x^0 = 0.52$.
In fact, as discussed below, this difference in the convergence rate 
of $\lambda^S(i)$ is relevant to the BCD of the downstream dynamics.

Since we have assumed that the fixed point pattern is AS in the downstream,
the noise has to be amplified
at a lattice point where the fixed point remains CU, 
in order for NSS to be formed.
First, we estimate the lattice point $i_u$, defined as the site
where  the convective instability is lost; 
in other words,
the site where the fixed point $x^i_*$ changes from CU to AS.  
Recall that the $x^i_*$ approach a periodic pattern in the downstream 
in the absence of noise.  
Denoting this spatial period by $L$,
the lattice point $i_u$ is given by the point $i$ such that
$\sum_{\ell =0}^{L-1} \lambda^S(i-\ell) / L$
is positive for $i < i_u$ and negative for $i \geq i_u$.
As can be also expected by considering Figs.\ref{fig:i-fix} and Fig.\ref{fig:i-bun},  
$i_u$ is strongly correlated with the relaxation scale of the
spatial map.  If the convergence of the spatial map to its attractor
is more rapid, $i_u$ is generally smaller. 
(See Fig.1: for example, $i_u = 12$ for $x^0 = 0.60$ indicated by $\bullet$, 
while $i_u = 36$ for $x^0 = 0.52$ indicated by $\times$.)

On the other hand, the scale $i_g$  required for the amplification of
a tiny noise to $O(1)$ can be estimated by
\begin{math}
\sigma \mbox{exp} (\sum^{i_g}_{i=1}\lambda^S(i)) \sim 1.
\end{math}
Then, the condition \cite{Fuji-KK} for the formation of NSS is simply given by
\begin{equation}
i_u - i_g  \geq 0.
\end{equation}

\begin{figure}
\begin{center}
\includegraphics[scale=0.55]{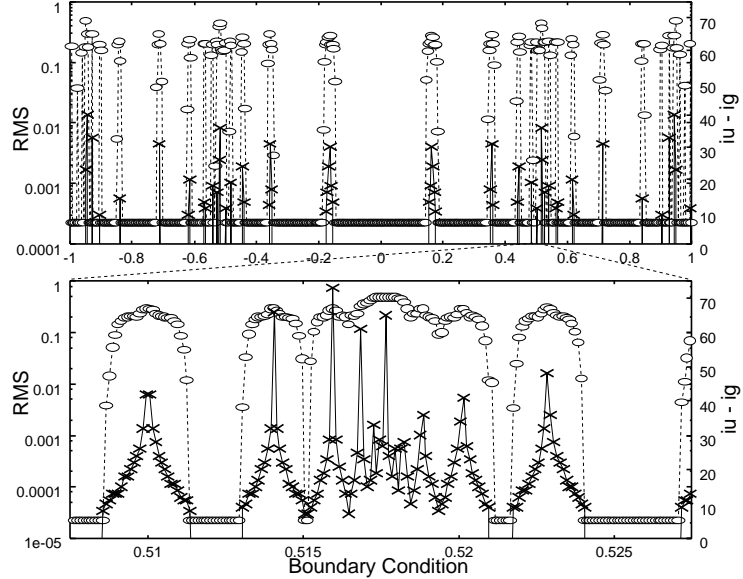}
\caption{BCD of the RMS of $x^{200}$ ($\circ$) and 
$i_u-i_g$ ($\times$), with $\epsilon = 0.9108$, spatial period = 4, 
and $\sigma =  2 \times 10^{-4}$.  
The lower figure is the blow-up of the upper figure.}
\label{fig:SP4}
\end{center}
\end{figure}

\begin{figure}
\begin{center}
\includegraphics[scale=0.60]{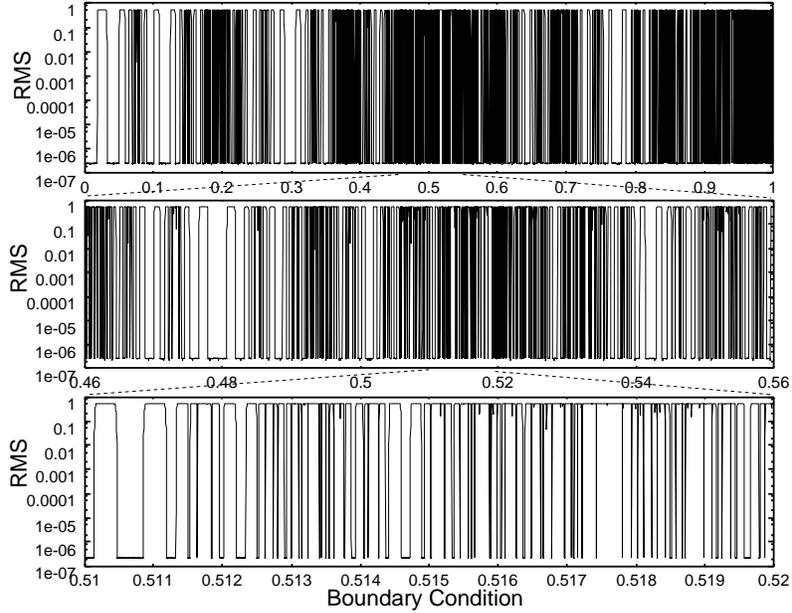}
\caption{BCD of the RMS  of $x^{200}$ with  
$\epsilon = 0.928198$, spatial period = 16, and $\sigma =  2 \times 10^{-8}$.
Successive blow-ups are given from the top to bottom figure.
Each figure represents 5000 data points taken at equal intervals.}
\label{fig:SP16}
\end{center}
\end{figure}

In Fig.\ref{fig:SP4}, we have plotted $i_u-i_g$ and the RMS of downstream 
dynamics as function of the boundary value $x^0$.  
The numerical results clearly support the conclusion that 
the condition for NSS is given by $i_u - i_g  \geq 0$.  
Although this condition concerns only  the sign of  $i_u - i_g$, 
the amplitude of the NSS is also highly correlated
with the value of $i_u - i_g$, as shown in Fig.\ref{fig:SP4}.
In the case of  Fig.\ref{fig:i-bun}, the downstream 
AS dynamics are of spatial period 2, where
NSS appears around $x^0 = 0.52$ and $0.97$. 
The BCD here is simple, with just two regions allowing for NSS.  
For the case of spatial period 4, as shown in
Fig.\ref{fig:SP4}, there are many undulation, and this BCD
has a self-similar fine structure, to be shown as fractals (see the blow-up of Fig.\ref{fig:SP4}).
The complexity of this self-similar structure of the  BCD increases 
with the period, as shown in Fig.\ref{fig:SP16}
for the case  of spatial period 16.
With this self-similar structure, a small difference in the boundary 
value  results in a large difference in the downstream dynamics.

We now discuss the origin of such sensitive BCD.
As seen from Figs.\ref{fig:SP4} and \ref{fig:SP16}, 
this BCD is due to the complicated structure of the BCD of  $i_u - i_g$.
Here, $i_g$ has a rather smooth dependence on $x^0$, and
the complicated structure is  due mainly to $i_u$.  In fact, there are (infinitely) many
local maxima of $i_u$ considered as a function of $x^0$,
in analogy to the plot of $i_u - i_g$ in Fig.\ref{fig:SP4}.

Recall that the scale
$i_u$ is highly correlated with the duration of the transient process of the
spatial map Eq.(\ref{eq:spacemap}), i.e., the number of steps required
for an orbit, generated by the map starting from $x^0$, to fall into
a periodic attractor.
When the period of the attractor of the map $g(x)$ is $L$, 
there are $L$ stable fixed points for the map  $x \rightarrow g^L(x)$.
Each stable fixed point corresponds to a different 
phase of the periodic attractor of the spatial map $g(x)$.  
For each value of the initial condition $x^0$ in the spatial map, 
the fixed point to which the $g^L(x)$ map is attracted 
[i.e., the phase of the cycle in the map $g(x)$] is different.

\begin{figure}
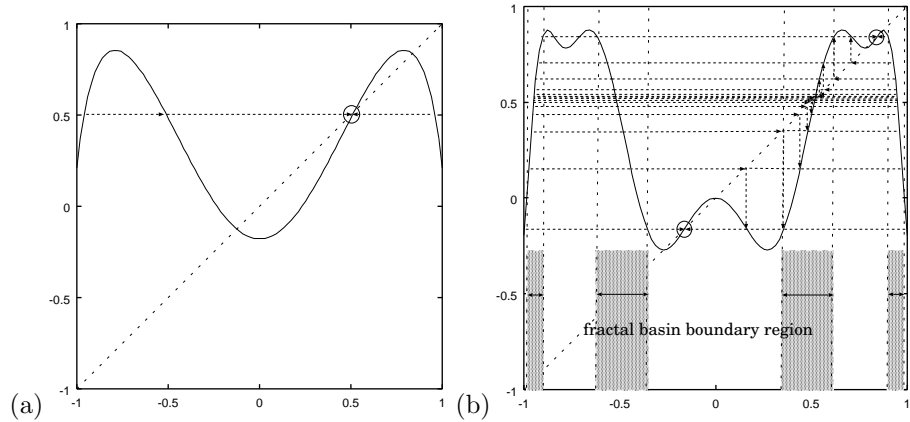

\begin{center}
(a)\includegraphics[scale=0.45]{Fig6a.eps}
(b)\includegraphics[scale=0.45]{Fig6b.eps}
\caption{The orbits falling onto unstable fixed point for
the spatial map $g^L(x)$.
All the points on the orbits are the basin boundary points,
where the phase of the spatial cycle shows slips.  
There are 4 such points in the spatial 
period 2 case (a), and an  infinite number in the spatial period 4 case (b).}
\label{fig:spacemap}
\end{center}
\end{figure}

When multiple attractors coexist, 
it is often the case that the basin structure
for each attractor is fractal \cite{GOY}. 
This is true for $g^L(x)$, and here 
there can be infinitely many basin boundary points.  
For example, in Fig.\ref{fig:spacemap} we have plotted the basin boundary points in the spatial map.
For the spatial period 2 case, there are 4 such points, as shown in
Fig.\ref{fig:spacemap}(a), while there are infinitely many points and the basin boundary is fractal
for the case of spatial period 4 (or higher), as shown in Fig.\ref{fig:spacemap}(b). 
Note that at unstable fixed points of $g^L(x)$, 
the phase of the attracted cycle slips.  
Successive preimages of such fixed points are nothing but the basin 
boundary of the map  $g^L(x)$, with which
fractal basin boundary is formed.  

For each basin boundary point, the number of transient steps before the attraction of 
an orbit to an attractor diverges. 
Hence, $i_u$ takes a local maximum at each point.
Accordingly, there are infinitely many local maxima of $i_u$, organized in
a self-similar manner.  
Now, the sensitive dependence on  $x^0$ is 
understood resulting from  a fractal basin boundary in the spatial map.
Since this formation of a fractal basin boundary
is rather common in  one-dimensional maps (with topological chaos),
sensitive BCD is expected to be a general phenomenon.
Here, it should be stressed that although the mechanism here is based on the 
fractal basin in the spatial map, the sensitive dependence is 
explained as a dependence on 
the {\sl boundary condition} rather than the initial condition\cite{Comm2}.

In conclusion, we have demonstrated the existence of a sensitive BCD
in a system characterized by convective instability in the upstream.  
Note that the loss of convective instability in the downstream 
(for the noiseless case) is
necessary to have such BCD.
If the downstream without noise is CU, 
then the BCD becomes weaker as the distance from the boundary increases,
and in an infinitely large system, this BCD eventually dies
away completely.

The origin of boundary condition sensitivity is found to be  the complex 
transient dynamics in the spatial map,
which lead to a fractal basin boundary.   
Accordingly, a sensitive BCD is generally expected as long
as the spatial map exhibits topological (or transient) chaos.
The analysis presented here can be extended to the case in which, 
in the absence of noise, the downstream does not possess fixed points,
but, rather possesses a stable cycle.

Although we have studied the sensitive BCD for the simplest case with a CML, 
our analysis can be straightforwardly extended to systems of  ODE.  
Convective instability in open flow is observed in  a wide variety of systems
with fluctuations, including  chemical 
reaction networks\cite{Fuji-KK,Rovinsky-Menzinger}, 
optical networks\cite{Ikeda-Otsuka}, traffic flow,
and open fluid flow.
Sensitive BCD is expected to be
observed in an such systems.
In particular, sensitive BCD in chemical reaction networks may be important 
in understanding diverse responses in  signal transduction systems of cells.

This work is partially supported by Grants-in-Aid for Scientific Research
from the Ministry of Education, Science, and Culture of Japan (11CE2006 and 11837004).

\begin{thebibliography}{99}
\bibitem{Comm0}
Here we apply the noise to all elements. However, the noise only to the 
upstream is essential. In fact, we obtain the same results, even for
a system with a noise only at the boundary $i = 1$.
\bibitem{Comm2}
Of course, dependence on the boundary value with a finer scale
than the applied noise is blurred.
Since the noise has to exist to establish a finite $i_g$,
the self-similar structure is seen only down to $\sigma$, (which is typically 
very small).
\end{thebibliography}
\end{document}